\documentclass[aps,amsmath,preprintnumbers,amssymb,twocolumn,superscriptaddress,nofootinbib,longbibliography,showpacs]{revtex4-1}

\usepackage{graphicx}
\usepackage{dcolumn}
\usepackage{bm}
\usepackage{color}
\usepackage{comment}
\usepackage{float}
\usepackage[normalem]{ulem}
\usepackage[colorlinks=true,pdfstartview=FitV,linkcolor=blue,citecolor=blue,urlcolor=blue]{hyperref}
\usepackage[mathlines]{lineno}

\everymath{\displaystyle}
\begin{document}

\newcommand{\up}[1]{$^{#1}$}
\newcommand{\down}[1]{$_{#1}$}
\newcommand{\powero}[1]{\mbox{10$^{#1}$}}
\newcommand{\powert}[2]{\mbox{#2$\times$10$^{#1}$}}

\newcommand{\mchi}{\mbox{$m_\chi$}}
\newcommand{\dedx}{$\rm{dE/dx}$}
\newcommand{\gev}{\mbox{GeV\,}$c^{-2}$}
\newcommand{\swn}{\mbox{$\sigma_{\chi-n}$}}
\newcommand{\evr}{\mbox{eV$_{\rm nr}$}}
\newcommand{\eve}{\mbox{eV$_{\rm ee}$}}
\newcommand{\dru}{\mbox{keV$_{\rm ee}^{-1}$\,kg$^{-1}$\,day$^{-1}$}}
\newcommand{\um}{\mbox{$\mu$m}}
\newcommand{\sxy}{\mbox{$\sigma_{xy}$}}
\newcommand{\sx}{\mbox{$\sigma_{x}$}}
\newcommand{\smax}{\mbox{$\sigma_{\rm max}$}}
\newcommand{\spix}{\mbox{$\sigma_{\rm pix}$}}
\newcommand{\obo}{\mbox{1$\times$1}}
\newcommand{\obh}{\mbox{1$\times$100}}
\newcommand{\dll}{\mbox{$\Delta LL$}}
\newcommand*\diff{\mathop{}\!\mathrm{d}}
\newcommand*\Diff[1]{\mathop{}\!\mathrm{d^#1}}

\newcommand{\kgd}{kg-day}
\newcommand{\tritium}{\mbox{$^{3}$H}}
\newcommand{\ironfive}{\mbox{$^{55}$Fe}}
\newcommand{\coseven}{\mbox{$^{57}$Co}}
\newcommand{\pbten}{$^{210}$Pb}
\newcommand{\biten}{$^{210}$Bi}
\newcommand{\sitwo}{$^{32}$Si}
\newcommand{\ptwo}{$^{32}$P}
\newcommand{\Ez}{\mbox{$E\text{-}z$} }
\newcommand{\Esigma}{\mbox{$E\text{-}\sigma_x$} }

\newcommand{\todo}[1]{\textcolor{red}{#1}}

\title{The DAMIC excess from WIMP-nucleus elastic scattering}

\author{A.E.\,Chavarria}
\affiliation{Center for Experimental Nuclear Physics and Astrophysics, University of Washington, Seattle, WA, United States}

\author{H.\,Lin}
\affiliation{Center for Experimental Nuclear Physics and Astrophysics, University of Washington, Seattle, WA, United States}

\author{K.J.\,McGuire}
\affiliation{Center for Experimental Nuclear Physics and Astrophysics, University of Washington, Seattle, WA, United States}

\author{A.\,Piers}
\affiliation{Center for Experimental Nuclear Physics and Astrophysics, University of Washington, Seattle, WA, United States}

\author{M.\,Traina}
\affiliation{Center for Experimental Nuclear Physics and Astrophysics, University of Washington, Seattle, WA, United States}

\noaffiliation

\date{\today}

\begin{abstract}
Two dark matter searches performed with charge-coupled devices (CCDs) in the DAMIC cryostat at SNOLAB reported with high statistical significance the presence of an unidentified source of low-energy events in bulk silicon.
The observed spectrum is consistent with nuclear recoils from the elastic scattering of weakly interacting massive particles (WIMPs) with masses between 2 and 4\,\gev. 
In the standard scenario of spin-independent WIMP-nucleus scattering, the derived cross section is conclusively excluded by results in argon by the DarkSide-50 experiment.
We identify isospin-violating and spin-dependent scenarios where interactions with $^{40}$Ar are strongly suppressed and the interpretation of the DAMIC excess as WIMP-nucleus elastic scattering remains viable.
\end{abstract}


\maketitle

Particle dark matter is one of the most exciting puzzles in contemporary physics and cosmology~\cite{Kolb:1990vq}.
Despite comprising around a quarter of the universe's total mass-energy budget, its elusive nature remains enigmatic.
Several particle candidates have been proposed to constitute the dark matter.
Of these, weakly interacting massive particles (WIMPs) remain one of the most compelling~\cite{Griest:2000kj, *Zurek:2013wia} and are actively searched for by highly sensitive detectors deployed in low-radiation environments deep underground.
The expected signal from WIMP interactions is the recoiling atom with $\mathcal{O}$(keV) of kinetic energy following WIMP-nucleus scattering in the detection target.
Throughout the past two decades, compelling hints of possible signals have been reported by DAMA in sodium iodide~\cite{Bernabei:2018jrt}, CoGeNT in germanium~\cite{CoGeNT:2012sne} and CDMS-II in silicon~\cite{CDMS:2013juh} but none so far have been confirmed, and all are in conflict with results from other experiments.
Notoriously, liquid xenon detectors have improved in sensitivity by several orders of magnitude beyond where these signals are expected, reporting consistently null results~\cite{XENON:2019S2, *LUX:2019DPE, *XENON:2020B8, *PandaX:2022B8, *PandaX:2022S2, LZ:2022xrq, *XENONNT:2023orw}.
However, since the new physics that mediates the WIMP-nucleus interaction is unknown, comparisons between results in different nuclear targets require theoretical assumptions, and great effort has been devoted in the theory community to cover the full breadth of possibilities~\cite{Giuliani:2005my, *Feng:2011vu, Fitzpatrick:2012ix, Hoferichter:2018acd}.
Nevertheless, to minimize theoretical uncertainty, other experimental efforts have focused on reproducing the signals with the same targets, with ANAIS~\cite{ANAIS:2019jul} and SuperCDMS~\cite{SuperCDMS:2018PL} obtaining null results in sodium iodide and germanium, respectively.
All things considered, there is so far no conclusive evidence for the existence of WIMPs.

Recently, the DAMIC, DAMIC-M and SENSEI Collaborations reported with 3.4\,$\sigma$ statistical significance the presence of an unidentified source of low-energy ionization events in the bulk silicon of skipper charge-coupled devices (CCDs) deployed in the DAMIC cryostat at SNOLAB~\cite{DAMIC:2023ela}.
This result confirms the original observation of the excess population of events above the background model in a previous installation of the DAMIC detector in 2020~\cite{DAMIC:2020prl, DAMIC:2021prd}.
Although both experiments were performed in the same cryostat, they are significantly different in terms of the CCDs used, the readout strategy, and the data analysis.
The collaborations remain agnostic about the origin of the excess and suggest an unidentified source of neutrons as a possibility without discarding an astrophysical origin.
The main difficulty with the neutron hypothesis is that the required neutron flux of 0.2\,cm$^{-2}$\,day$^{-1}$ is $>$100 times larger than estimated from known background sources~\cite{DAMIC:2021prd} and comparable to the total neutron flux in the SNOLAB cavern~\cite{snolabuh}.
Thus, the unidentified source inside the DAMIC detector's 42-cm thick polyethylene shield would need to be so strong as to build up a flux inside the shield that is as large as in the cavern outside.
Since there are no straightforward known-physics explanations for the DAMIC excess, we consider astrophysical interpretations from interactions of the expected flux of dark matter particles through the DAMIC detector.

\begin{figure}[b!]
	\centering
	\includegraphics[width=0.48\textwidth]{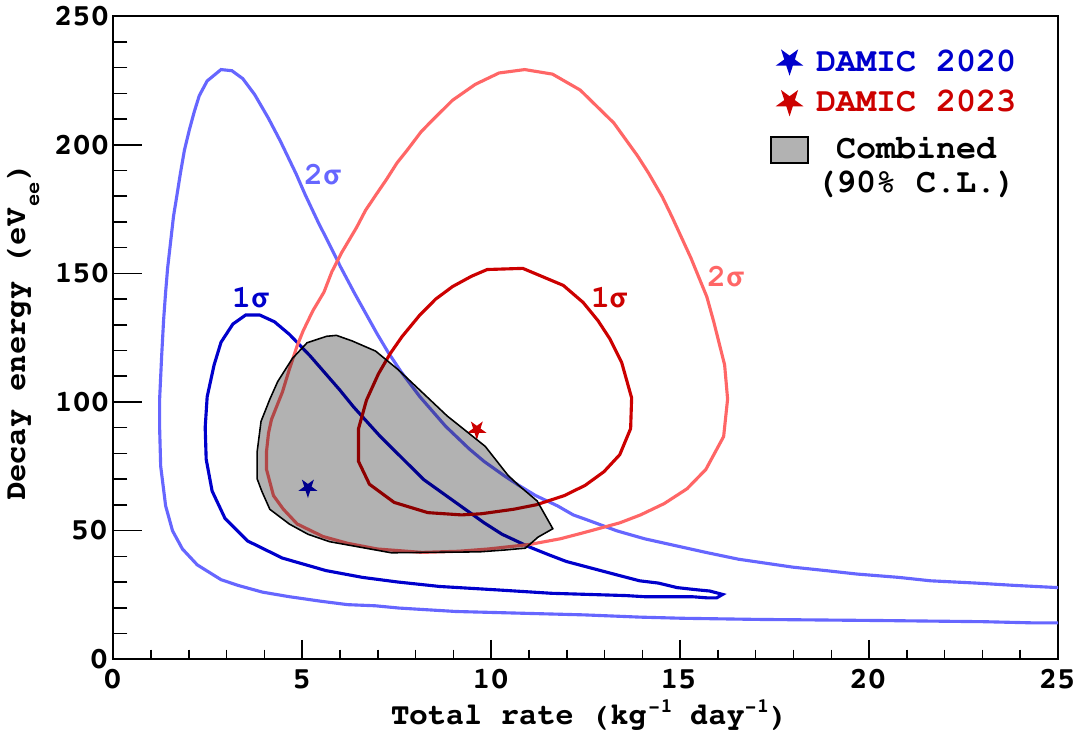}
	\caption{Derived 90\% C.L. signal region in $(R,\epsilon)$ space after combining the results from the two DAMIC experiments. The combined region is overlaid on the contours from Ref.~\cite{DAMIC:2023ela}. }
	\label{fig:contours}
\end{figure}

\begin{figure}[t!]
	\centering
	\includegraphics[width=0.48\textwidth]{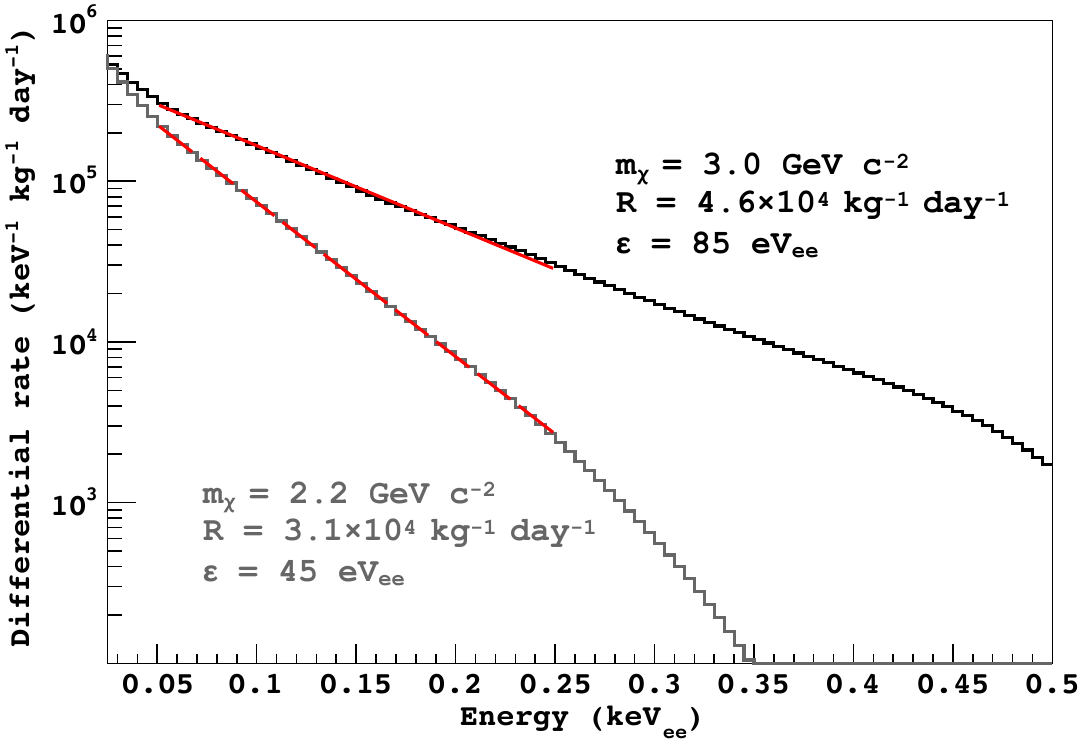}
	\caption{Calculated WIMP signal spectra in silicon for $\sigma_N$$=$1\,pb and two values of $m_\chi$ following the procedure described in the text. The spectra are fit between 50 and 250\,\eve\ to obtain $R$ and $\epsilon$ of the exponential approximation with the best-fit function in red and corresponding parameters in the legend.}
	\label{fig:spec}
\end{figure}

\begin{figure}[t!]
	\centering
	\includegraphics[width=0.48\textwidth]{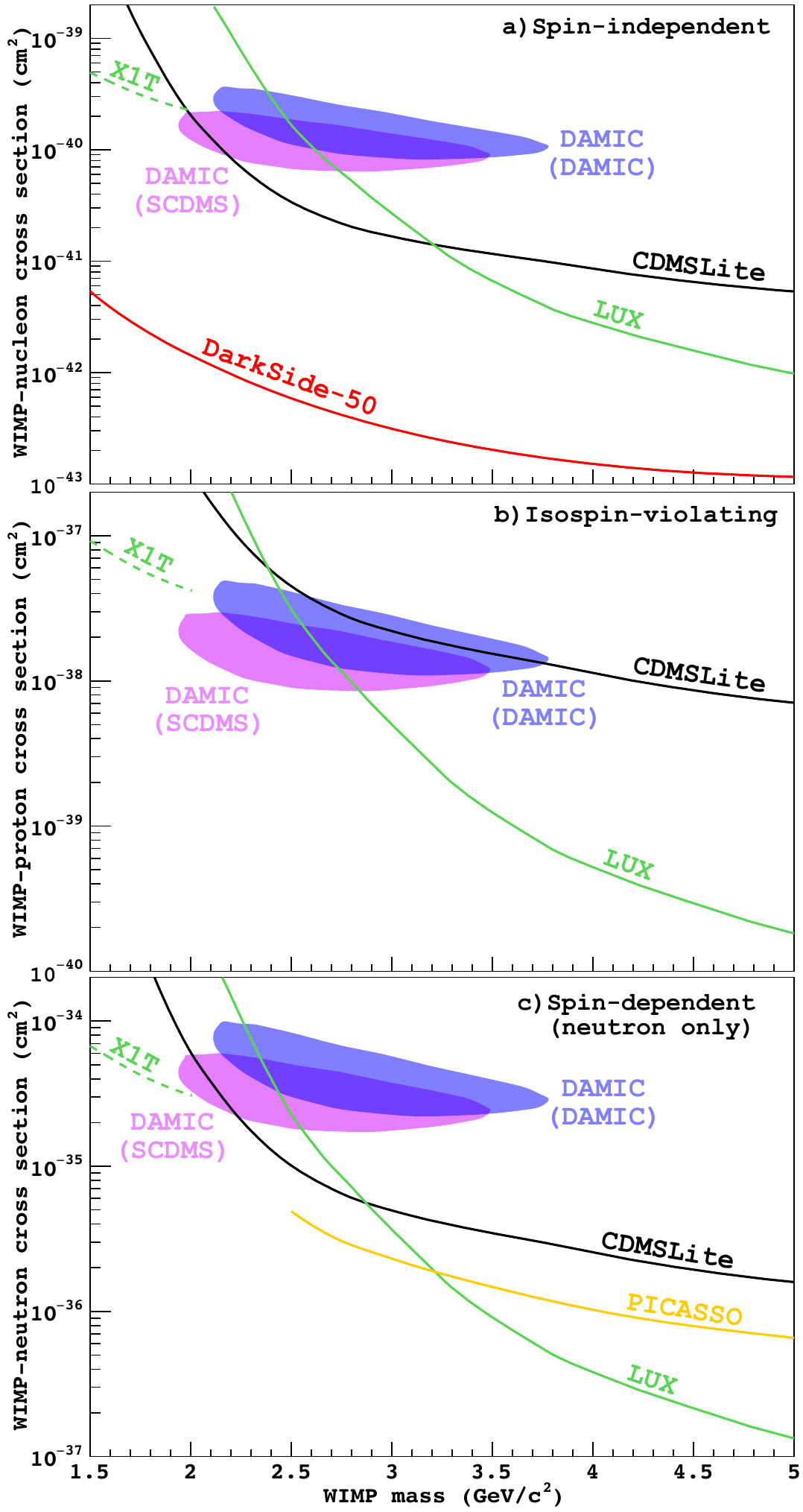}
	\caption{Signal region (90\% C.L.) corresponding to the DAMIC spectral excess for three models of WIMP-nucleus elastic scattering. The shaded blue (pink) region assumes the nuclear-recoil ionization efficiency measurements by DAMIC~\cite{Chavarria:2016xsi} (SuperCDMS~\cite{SuperCDMS:2023geu}). Exclusion limits (90\% C.L.) from DarkSide-50~\cite{DarkSide-50:2022qzh} (solid red), CDMSLite~\cite{SuperCDMS:2018PL} (solid black), LUX~\cite{LUX:2020S2} (solid green), PICASSO~\cite{PICASSO:2016lsk} (solid yellow), and XENON1T~\cite{XENON:2019migdal} (dashed green) are also shown. a)~Standard spin-independent scenario, b)~Isospin-violating scenario with $f_n/f_p$$=$$-0.82$, c)~Spin-dependent scenario with $a_p/a_n$$=$0.}
	\label{fig:excl}
\end{figure}

The DAMIC detector measures the spectrum of ionization events in electron-equivalent energy units (\eve ), which correspond to the energy of electronic recoils that would result in the observed ionization signals.
The DAMIC excess is well parametrized by an exponentially decaying spectrum, which, above DAMIC's 23\,\eve\ analysis threshold, is a good approximation to the spectrum from the elastic scattering of WIMPs with silicon nuclei.
Reference~\cite{DAMIC:2023ela} provides the allowed parameter space for the decaying exponential in terms of the total rate $R$ ($\sim$7 events per kg-day) and the decay energy $\epsilon$ ($\sim$80\,\eve ) with 1 and 2\,$\sigma$ contours for both the 2020 and 2024 results.
We start from the best-fit value of each result and perform a linear interpolation along the radial direction between the contours.
We find points on the parameter space with combined $p$ value of 0.1  to construct the allowed region with 90\% C.L. shown in Fig.~\ref{fig:contours}.
Each pair of values $(R,\epsilon)$ uniquely corresponds to the signal spectrum from a WIMP with spin-independent WIMP-nucleon scattering cross section $\sigma_N$ and mass $m_{\chi}$.
We generate nuclear-recoil spectra for different values of $m_\chi$ and a reference value of $\sigma_N$$=$$1$\,pb$=$$10^{-36}$\,cm$^{-2}$ following the prescription from Ref.~\cite{Lewin:1995rx} with WIMP speed distribution from Ref.~\cite{Baxter:2021pqo}.
Since nuclear recoils generate a smaller ionization signal than an electronic recoil of the same energy, we convert the resulting nuclear-recoil energy spectrum to electron-equivalent energies using the nuclear-recoil ionization efficiency model for silicon obtained from CCD calibrations~\cite{Chavarria:2016xsi} with the parametrization from Ref.~\cite{SuperCDMS:2016proj}.
To model the detector energy response, we assume a Fano factor of 1.
Finally, we fit the signal spectra with an exponential function from 50 to 250\,\eve\ to extract the corresponding $(R,\epsilon)$.
Figure~\ref{fig:spec} shows the generated signal spectra for two values of $m_\chi$ and the corresponding best-fit exponential.
We use the fit results to translate the points along the contour of the combined 90\% C.L. region from $(R,\epsilon)$ space to $(\sigma_N,m_\chi)$ space.
The shaded blue region in Fig.~\ref{fig:excl}a shows the 90\% C.L. signal region that corresponds to the DAMIC excess.
The nuclear-recoil ionization efficiency is the most relevant systematic uncertainty in this extraction, especially since a recent measurement by the SuperCDMS Collaboration~\cite{SuperCDMS:2023geu} reports a significantly higher ionization efficiency than previously measured with CCDs.
Thus, we also present the 90\% C.L. signal region assuming the recent SuperCDMS measurement as the shaded region in pink.
We also show 90\% C.L. exclusion limits from experiments that exclude or are in the vicinity of the DAMIC region below 3\,\gev .
We omit for clarity other experiments that also exclude the DAMIC region above 3\,\gev~~\cite{PICO:2019vsc, *CRESST:2019jnq, XENON:2019S2, *LUX:2019DPE, *XENON:2020B8, *PandaX:2022B8, *PandaX:2022S2}, and assume that the parameter space is generally excluded above this mass.
The possible signals in CDMS-II Si would correspond to higher energy recoils from WIMPs with \mchi $>$5\,\gev\ and are unrelated to the DAMIC excess in this scenario.

Below 3\,\gev\ the DAMIC region is only conclusively excluded by DarkSide-50 (red line in Fig.~\ref{fig:excl}a). The DarkSide Collaboration measured ionization signals in a liquid argon time projection chamber (TPC) to probe the energy range of the DAMIC excess with sensitivity to interaction rates 50 times smaller than the rate observed in DAMIC and did not find any excess events above their background model.
Since we have not identified any systematic uncertainty in detector response or WIMP speed distribution that could resolve such strong disagreement, we turn to scenarios where the interaction of WIMPs with $^{40}$Ar (99.6\% abundance in natural argon~\cite{Kondev:2021lzi}) is suppressed.
We consider the first-order formalism~\cite{Kurylov:2003ra} where the new-physics details of the interaction between the WIMP and the nucleus is determined by four parameters: $f_{p,n}$ and $a_{p,n}$ for the coupling of the WIMP to the proton and neutron in spin-independent and spin-dependent scattering, respectively.
The ``standard'' spin-independent scenario presented above corresponds to $f_p$$=$$f_n$ and $a_p$$=$$a_n$$=$0.
In this case, the quoted WIMP-nucleon scattering cross $\sigma_N$ is equal to the spin-independent WIMP-proton scattering cross section $\sigma_p$.

Consider the isospin-violating spin-independent scenario with $f_n/f_p$$\neq$1~\cite{Giuliani:2005my, *Feng:2011vu}.
For certain values of $f_n/f_p$, interference can suppress the scattering cross section, vanishing in $^{40}$Ar for $f_n/f_p$$=$$-0.82$.
Reference~\cite{Feng:2013Snowmass} provides the ``degradation factor'' $\sigma_N / \sigma_p$ for different targets as a function of $f_n/f_p$, which we use to recast the DAMIC region and other exclusion limits in Fig.~\ref{fig:excl}b.
For reference, the CDMSLite result used germanium as the target, while LUX and XENON1T used xenon.
In this scenario, a region of the DAMIC excess parameter space between 2 and 3\,\gev\ remains unexcluded and constitutes a possible dark matter origin of the DAMIC excess.
Note that in the interference regime next-to-leading-order (NLO) corrections dominate the scattering cross section~\cite{Cirigliano:2012pq, *Cirigliano:2013zta} and total suppression in $^{40}$Ar may not occur at $f_n/f_p$$=$$-0.82$.
More complete treatments of spin-independent WIMP-nucleus scattering result in a larger number of free model parameters~\cite{Hoferichter:2018acd, Cirigliano:2012pq, *Cirigliano:2013zta}, which can also be suitably chosen for cancellation to occur in $^{40}$Ar and realize scenarios similar to Fig.~\ref{fig:excl}b.

Since $^{40}$Ar does not have a nuclear spin, interactions from WIMPs that couple to the nuclear spin are suppressed in $^{40}$Ar.
While silicon is 92.2\% spinless $^{28}$Si, it has a 4.7\% isotopic abundance of $^{29}$Si with nuclear spin $J$$=$1/2~\cite{Kondev:2021lzi}, which provides DAMIC with some sensitivity to spin-dependent interactions.
Germanium and xenon targets also have isotopes with non-zero spin.
Reference~\cite{Buckley:2013gjo} provides example calculations in the first order-formalism of spin-dependent cross sections for different targets, where results are provided for cases where the WIMP only couples to the proton ($a_n/a_p$$=$0) or to the neutron ($a_p/a_n$$=$0).
Note that although $^{40}$Ar is insensitive to WIMPs at tree level, several effective-field theory (EFT) studies show that, through operator mixing effects, models where spin-dependent scattering is dominant often predict non-negligible scattering rates with spinless nuclei~\cite{Crivellin:2014qxa, *Bishara:2018vix, *Cheek:2023zhv}.
Notwithstanding, we use Ref.~\cite{Buckley:2013gjo} to recast the DAMIC region and other experiments' exclusion limits in the spin-dependent parameter space, which is shown in Fig.~\ref{fig:excl}c for the case where the WIMP couples only to the neutron.
We confirm that the recast limits are consistent with published spin-dependent limits where available~\cite{SuperCDMS:2017SD, XENON:2019migdal}.
We include results from the PICASSO experiment~\cite{PICASSO:2016lsk}, which uses C$_4$F$_{10}$ as a target and excludes a fraction of the DAMIC region.
In this scenario, there is a small fraction of the the DAMIC region between 2.0 and 2.2 \gev\ that is not excluded by any published limit.
Extrapolations of the XENON1T and PICASSO limits could be in tension with this interpretation of the DAMIC excess.
However, we note that extrapolations of the PICASSO limit toward lower masses suffer from significant systematic uncertainties in the WIMP speed distribution and the response of the detector at the lowest energies, while the XENON1T limit relies on the Migdal effect~\cite{Ibe:2017yqa}, an atomic process that has never been observed in the laboratory at the energies relevant for dark matter searches~\cite{Xu:2023wev}.
For the case where the spin-dependent interaction couples to the proton, the exclusion limits from the other experiments reach lower cross sections relative to the DAMIC region.
Thus, the explanation for the DAMIC excess by spin-dependent WIMP-nucleus elastic scattering can more easily be realized if the WIMP couples mostly to the neutron.

The origin of the observed excess population of bulk events in DAMIC at SNOLAB remains a mystery.
The result has been reproduced with high statistical significance in two separate experiments.
In this letter, we considered the possibility that the event excess in DAMIC arises from the elastic scattering of WIMPs with $m_\chi$ between 2 and 4\,\gev , which induce silicon nuclear recoils with the observed spectrum.
The null result from the DarkSide-50 experiment requires scenarios where the WIMP interaction cross section with $^{40}$Ar is strongly suppressed.
We considered two such scenarios in the first-order formalism of WIMP scattering and find that regions of paramater space that correspond to the DAMIC excess remain viable.
These are isospin-violating spin-independent scattering where the proton and neutron couplings are chosen for maximum interference in $^{40}$Ar, and spin-dependent scattering where the WIMP couples only to the neutron.
In both cases, WIMPs with $m_\chi$ closer to 2.0\,\gev\ are favored, and more parameter space is allowed if the nuclear-recoil ionization efficiency is as recently measured by the SuperCDMS Collaboration.
Although this first-order study should be expanded to provide a more accurate and complete picture of the possible dark matter origin of the DAMIC excess, the conclusive resolution will come from the DAMIC-M~\cite{Privitera:2024tpq}, SENSEI~\cite{SENSEI:2020dpa} and SuperCDMS~\cite{SuperCDMS:2022kse} Collaborations, which will each operate more sensitive silicon detectors in the near future.

\begin{acknowledgments}
We are grateful to Vincenzo Cirigliano for insightful discussions on the limitations of the first-order formalism of WIMP-nucleus scattering.
We thank Ryan Roehnelt for useful comments on the manuscript.
The DAMIC group at the University of Washington is supported by the United States National Science Foundation (NSF) through Grant No.\ NSF PHY-2110585.
\end{acknowledgments}

\bibliography{myrefs.bib}

\begin{thebibliography}{49}%
\makeatletter
\providecommand \@ifxundefined [1]{%
 \@ifx{#1\undefined}
}%
\providecommand \@ifnum [1]{%
 \ifnum #1\expandafter \@firstoftwo
 \else \expandafter \@secondoftwo
 \fi
}%
\providecommand \@ifx [1]{%
 \ifx #1\expandafter \@firstoftwo
 \else \expandafter \@secondoftwo
 \fi
}%
\providecommand \natexlab [1]{#1}%
\providecommand \enquote  [1]{``#1''}%
\providecommand \bibnamefont  [1]{#1}%
\providecommand \bibfnamefont [1]{#1}%
\providecommand \citenamefont [1]{#1}%
\providecommand \href@noop [0]{\@secondoftwo}%
\providecommand \href [0]{\begingroup \@sanitize@url \@href}%
\providecommand \@href[1]{\@@startlink{#1}\@@href}%
\providecommand \@@href[1]{\endgroup#1\@@endlink}%
\providecommand \@sanitize@url [0]{\catcode `\\12\catcode `\$12\catcode
  `\&12\catcode `\#12\catcode `\^12\catcode `\_12\catcode `\%12\relax}%
\providecommand \@@startlink[1]{}%
\providecommand \@@endlink[0]{}%
\providecommand \url  [0]{\begingroup\@sanitize@url \@url }%
\providecommand \@url [1]{\endgroup\@href {#1}{\urlprefix }}%
\providecommand \urlprefix  [0]{URL }%
\providecommand \Eprint [0]{\href }%
\providecommand \doibase [0]{http://dx.doi.org/}%
\providecommand \selectlanguage [0]{\@gobble}%
\providecommand \bibinfo  [0]{\@secondoftwo}%
\providecommand \bibfield  [0]{\@secondoftwo}%
\providecommand \translation [1]{[#1]}%
\providecommand \BibitemOpen [0]{}%
\providecommand \bibitemStop [0]{}%
\providecommand \bibitemNoStop [0]{.\EOS\space}%
\providecommand \EOS [0]{\spacefactor3000\relax}%
\providecommand \BibitemShut  [1]{\csname bibitem#1\endcsname}%
\let\auto@bib@innerbib\@empty
\bibitem [{\citenamefont {Kolb}\ and\ \citenamefont
  {Turner}(1990)}]{Kolb:1990vq}%
  \BibitemOpen
  \bibfield  {author} {\bibinfo {author} {\bibfnamefont {Edward~W.}\
  \bibnamefont {Kolb}}\ and\ \bibinfo {author} {\bibfnamefont {Michael~S.}\
  \bibnamefont {Turner}},\ }\href@noop {} {\emph {\bibinfo {title} {{The Early
  Universe}}}},\ Vol.~\bibinfo {volume} {69}\ (\bibinfo  {publisher} {CRC
  Press},\ \bibinfo {year} {1990})\BibitemShut {NoStop}%
\bibitem [{\citenamefont {Griest}\ and\ \citenamefont
  {Kamionkowski}(2000)}]{Griest:2000kj}%
  \BibitemOpen
  \bibfield  {author} {\bibinfo {author} {\bibfnamefont {K.}~\bibnamefont
  {Griest}}\ and\ \bibinfo {author} {\bibfnamefont {M.}~\bibnamefont
  {Kamionkowski}},\ }\bibfield  {title} {\enquote {\bibinfo {title}
  {{Supersymmetric dark matter}},}\ }\href {\doibase
  10.1016/S0370-1573(00)00021-1} {\bibfield  {journal} {\bibinfo  {journal}
  {Phys. Rept.}\ }\textbf {\bibinfo {volume} {333}},\ \bibinfo {pages}
  {167--182} (\bibinfo {year} {2000})}\BibitemShut {NoStop}%
\bibitem [{\citenamefont {Zurek}(2014)}]{Zurek:2013wia}%
  \BibitemOpen
  \bibfield  {author} {\bibinfo {author} {\bibfnamefont {Kathryn~M.}\
  \bibnamefont {Zurek}},\ }\bibfield  {title} {\enquote {\bibinfo {title}
  {{Asymmetric Dark Matter: Theories, Signatures, and Constraints}},}\ }\href
  {\doibase 10.1016/j.physrep.2013.12.001} {\bibfield  {journal} {\bibinfo
  {journal} {Phys. Rept.}\ }\textbf {\bibinfo {volume} {537}},\ \bibinfo
  {pages} {91--121} (\bibinfo {year} {2014})},\ \Eprint
  {http://arxiv.org/abs/1308.0338} {arXiv:1308.0338 [hep-ph]} \BibitemShut
  {NoStop}%
\bibitem [{\citenamefont {Bernabei}\ \emph {et~al.}(2018)\citenamefont
  {Bernabei} \emph {et~al.}}]{Bernabei:2018jrt}%
  \BibitemOpen
  \bibfield  {author} {\bibinfo {author} {\bibfnamefont {R.}~\bibnamefont
  {Bernabei}} \emph {et~al.},\ }\bibfield  {title} {\enquote {\bibinfo {title}
  {{First model independent results from DAMA/LIBRA-phase2}},}\ }\href
  {\doibase 10.15407/jnpae2018.04.307} {\bibfield  {journal} {\bibinfo
  {journal} {Nucl. Phys. Atom. Energy}\ }\textbf {\bibinfo {volume} {19}},\
  \bibinfo {pages} {307--325} (\bibinfo {year} {2018})},\ \Eprint
  {http://arxiv.org/abs/1805.10486} {arXiv:1805.10486 [hep-ex]} \BibitemShut
  {NoStop}%
\bibitem [{\citenamefont {Aalseth}\ \emph {et~al.}(2013)\citenamefont {Aalseth}
  \emph {et~al.}}]{CoGeNT:2012sne}%
  \BibitemOpen
  \bibfield  {author} {\bibinfo {author} {\bibfnamefont {C.~E.}\ \bibnamefont
  {Aalseth}} \emph {et~al.} (\bibinfo {collaboration} {CoGeNT}),\ }\bibfield
  {title} {\enquote {\bibinfo {title} {{CoGeNT: A Search for Low-Mass Dark
  Matter using p-type Point Contact Germanium Detectors}},}\ }\href {\doibase
  10.1103/PhysRevD.88.012002} {\bibfield  {journal} {\bibinfo  {journal} {Phys.
  Rev. D}\ }\textbf {\bibinfo {volume} {88}},\ \bibinfo {pages} {012002}
  (\bibinfo {year} {2013})},\ \Eprint {http://arxiv.org/abs/1208.5737}
  {arXiv:1208.5737 [astro-ph.CO]} \BibitemShut {NoStop}%
\bibitem [{\citenamefont {Agnese}\ \emph {et~al.}(2013)\citenamefont {Agnese}
  \emph {et~al.}}]{CDMS:2013juh}%
  \BibitemOpen
  \bibfield  {author} {\bibinfo {author} {\bibfnamefont {R.}~\bibnamefont
  {Agnese}} \emph {et~al.} (\bibinfo {collaboration} {CDMS}),\ }\bibfield
  {title} {\enquote {\bibinfo {title} {{Silicon Detector Dark Matter Results
  from the Final Exposure of CDMS II}},}\ }\href {\doibase
  10.1103/PhysRevLett.111.251301} {\bibfield  {journal} {\bibinfo  {journal}
  {Phys. Rev. Lett.}\ }\textbf {\bibinfo {volume} {111}},\ \bibinfo {pages}
  {251301} (\bibinfo {year} {2013})},\ \Eprint {http://arxiv.org/abs/1304.4279}
  {arXiv:1304.4279 [hep-ex]} \BibitemShut {NoStop}%
\bibitem [{\citenamefont {Aprile}\ \emph
  {et~al.}(2019{\natexlab{a}})\citenamefont {Aprile} \emph
  {et~al.}}]{XENON:2019S2}%
  \BibitemOpen
  \bibfield  {author} {\bibinfo {author} {\bibfnamefont {E.}~\bibnamefont
  {Aprile}} \emph {et~al.} (\bibinfo {collaboration} {XENON}),\ }\bibfield
  {title} {\enquote {\bibinfo {title} {{Light Dark Matter Search with
  Ionization Signals in XENON1T}},}\ }\href {\doibase
  10.1103/PhysRevLett.123.251801} {\bibfield  {journal} {\bibinfo  {journal}
  {Phys. Rev. Lett.}\ }\textbf {\bibinfo {volume} {123}},\ \bibinfo {pages}
  {251801} (\bibinfo {year} {2019}{\natexlab{a}})},\ \Eprint
  {http://arxiv.org/abs/1907.11485} {arXiv:1907.11485 [hep-ex]} \BibitemShut
  {NoStop}%
\bibitem [{\citenamefont {Akerib}\ \emph {et~al.}(2020)\citenamefont {Akerib}
  \emph {et~al.}}]{LUX:2019DPE}%
  \BibitemOpen
  \bibfield  {author} {\bibinfo {author} {\bibfnamefont {D.~S.}\ \bibnamefont
  {Akerib}} \emph {et~al.} (\bibinfo {collaboration} {LUX}),\ }\bibfield
  {title} {\enquote {\bibinfo {title} {{Extending light WIMP searches to single
  scintillation photons in LUX}},}\ }\href {\doibase
  10.1103/PhysRevD.101.042001} {\bibfield  {journal} {\bibinfo  {journal}
  {Phys. Rev. D}\ }\textbf {\bibinfo {volume} {101}},\ \bibinfo {pages}
  {042001} (\bibinfo {year} {2020})},\ \Eprint
  {http://arxiv.org/abs/1907.06272} {arXiv:1907.06272 [astro-ph.CO]}
  \BibitemShut {NoStop}%
\bibitem [{\citenamefont {Aprile}\ \emph {et~al.}(2021)\citenamefont {Aprile}
  \emph {et~al.}}]{XENON:2020B8}%
  \BibitemOpen
  \bibfield  {author} {\bibinfo {author} {\bibfnamefont {E.}~\bibnamefont
  {Aprile}} \emph {et~al.} (\bibinfo {collaboration} {XENON}),\ }\bibfield
  {title} {\enquote {\bibinfo {title} {{Search for Coherent Elastic Scattering
  of Solar $^8$B Neutrinos in the XENON1T Dark Matter Experiment}},}\ }\href
  {\doibase 10.1103/PhysRevLett.126.091301} {\bibfield  {journal} {\bibinfo
  {journal} {Phys. Rev. Lett.}\ }\textbf {\bibinfo {volume} {126}},\ \bibinfo
  {pages} {091301} (\bibinfo {year} {2021})},\ \Eprint
  {http://arxiv.org/abs/2012.02846} {arXiv:2012.02846 [hep-ex]} \BibitemShut
  {NoStop}%
\bibitem [{\citenamefont {Ma}\ \emph {et~al.}(2023)\citenamefont {Ma} \emph
  {et~al.}}]{PandaX:2022B8}%
  \BibitemOpen
  \bibfield  {author} {\bibinfo {author} {\bibfnamefont {Wenbo}\ \bibnamefont
  {Ma}} \emph {et~al.} (\bibinfo {collaboration} {PandaX}),\ }\bibfield
  {title} {\enquote {\bibinfo {title} {{Search for Solar B8 Neutrinos in the
  PandaX-4T Experiment Using Neutrino-Nucleus Coherent Scattering}},}\ }\href
  {\doibase 10.1103/PhysRevLett.130.021802} {\bibfield  {journal} {\bibinfo
  {journal} {Phys. Rev. Lett.}\ }\textbf {\bibinfo {volume} {130}},\ \bibinfo
  {pages} {021802} (\bibinfo {year} {2023})},\ \Eprint
  {http://arxiv.org/abs/2207.04883} {arXiv:2207.04883 [hep-ex]} \BibitemShut
  {NoStop}%
\bibitem [{\citenamefont {Li}\ \emph {et~al.}(2023)\citenamefont {Li} \emph
  {et~al.}}]{PandaX:2022S2}%
  \BibitemOpen
  \bibfield  {author} {\bibinfo {author} {\bibfnamefont {Shuaijie}\
  \bibnamefont {Li}} \emph {et~al.} (\bibinfo {collaboration} {PandaX}),\
  }\bibfield  {title} {\enquote {\bibinfo {title} {{Search for Light Dark
  Matter with Ionization Signals in the PandaX-4T Experiment}},}\ }\href
  {\doibase 10.1103/PhysRevLett.130.261001} {\bibfield  {journal} {\bibinfo
  {journal} {Phys. Rev. Lett.}\ }\textbf {\bibinfo {volume} {130}},\ \bibinfo
  {pages} {261001} (\bibinfo {year} {2023})},\ \Eprint
  {http://arxiv.org/abs/2212.10067} {arXiv:2212.10067 [hep-ex]} \BibitemShut
  {NoStop}%
\bibitem [{\citenamefont {Aalbers}\ \emph {et~al.}(2023)\citenamefont {Aalbers}
  \emph {et~al.}}]{LZ:2022xrq}%
  \BibitemOpen
  \bibfield  {author} {\bibinfo {author} {\bibfnamefont {J.}~\bibnamefont
  {Aalbers}} \emph {et~al.} (\bibinfo {collaboration} {LUX-ZEPLIN}),\
  }\bibfield  {title} {\enquote {\bibinfo {title} {{First Dark Matter Search
  Results from the LUX-ZEPLIN (LZ) Experiment}},}\ }\href {\doibase
  10.1103/PhysRevLett.131.041002} {\bibfield  {journal} {\bibinfo  {journal}
  {Phys. Rev. Lett.}\ }\textbf {\bibinfo {volume} {131}},\ \bibinfo {pages}
  {041002} (\bibinfo {year} {2023})},\ \Eprint
  {http://arxiv.org/abs/2207.03764} {arXiv:2207.03764 [hep-ex]} \BibitemShut
  {NoStop}%
\bibitem [{\citenamefont {Aprile}\ \emph {et~al.}(2023)\citenamefont {Aprile}
  \emph {et~al.}}]{XENONNT:2023orw}%
  \BibitemOpen
  \bibfield  {author} {\bibinfo {author} {\bibfnamefont {E.}~\bibnamefont
  {Aprile}} \emph {et~al.} (\bibinfo {collaboration} {XENON}),\ }\bibfield
  {title} {\enquote {\bibinfo {title} {{First Dark Matter Search with Nuclear
  Recoils from the XENONnT Experiment}},}\ }\href {\doibase
  10.1103/PhysRevLett.131.041003} {\bibfield  {journal} {\bibinfo  {journal}
  {Phys. Rev. Lett.}\ }\textbf {\bibinfo {volume} {131}},\ \bibinfo {pages}
  {041003} (\bibinfo {year} {2023})},\ \Eprint
  {http://arxiv.org/abs/2303.14729} {arXiv:2303.14729 [hep-ex]} \BibitemShut
  {NoStop}%
\bibitem [{\citenamefont {Giuliani}(2005)}]{Giuliani:2005my}%
  \BibitemOpen
  \bibfield  {author} {\bibinfo {author} {\bibfnamefont {F.}~\bibnamefont
  {Giuliani}},\ }\bibfield  {title} {\enquote {\bibinfo {title} {{Are direct
  search experiments sensitive to all spin-independent WIMP candidates?}}}\
  }\href {\doibase 10.1103/PhysRevLett.95.101301} {\bibfield  {journal}
  {\bibinfo  {journal} {Phys. Rev. Lett.}\ }\textbf {\bibinfo {volume} {95}},\
  \bibinfo {pages} {101301} (\bibinfo {year} {2005})},\ \Eprint
  {http://arxiv.org/abs/hep-ph/0504157} {arXiv:hep-ph/0504157} \BibitemShut
  {NoStop}%
\bibitem [{\citenamefont {Feng}\ \emph {et~al.}(2011)\citenamefont {Feng},
  \citenamefont {Kumar}, \citenamefont {Marfatia},\ and\ \citenamefont
  {Sanford}}]{Feng:2011vu}%
  \BibitemOpen
  \bibfield  {author} {\bibinfo {author} {\bibfnamefont {Jonathan~L.}\
  \bibnamefont {Feng}}, \bibinfo {author} {\bibfnamefont {Jason}\ \bibnamefont
  {Kumar}}, \bibinfo {author} {\bibfnamefont {Danny}\ \bibnamefont {Marfatia}},
  \ and\ \bibinfo {author} {\bibfnamefont {David}\ \bibnamefont {Sanford}},\
  }\bibfield  {title} {\enquote {\bibinfo {title} {{Isospin-Violating Dark
  Matter}},}\ }\href {\doibase 10.1016/j.physletb.2011.07.083} {\bibfield
  {journal} {\bibinfo  {journal} {Phys. Lett. B}\ }\textbf {\bibinfo {volume}
  {703}},\ \bibinfo {pages} {124--127} (\bibinfo {year} {2011})},\ \Eprint
  {http://arxiv.org/abs/1102.4331} {arXiv:1102.4331 [hep-ph]} \BibitemShut
  {NoStop}%
\bibitem [{\citenamefont {Fitzpatrick}\ \emph {et~al.}(2013)\citenamefont
  {Fitzpatrick}, \citenamefont {Haxton}, \citenamefont {Katz}, \citenamefont
  {Lubbers},\ and\ \citenamefont {Xu}}]{Fitzpatrick:2012ix}%
  \BibitemOpen
  \bibfield  {author} {\bibinfo {author} {\bibfnamefont {A.~Liam}\ \bibnamefont
  {Fitzpatrick}}, \bibinfo {author} {\bibfnamefont {Wick}\ \bibnamefont
  {Haxton}}, \bibinfo {author} {\bibfnamefont {Emanuel}\ \bibnamefont {Katz}},
  \bibinfo {author} {\bibfnamefont {Nicholas}\ \bibnamefont {Lubbers}}, \ and\
  \bibinfo {author} {\bibfnamefont {Yiming}\ \bibnamefont {Xu}},\ }\bibfield
  {title} {\enquote {\bibinfo {title} {{The Effective Field Theory of Dark
  Matter Direct Detection}},}\ }\href {\doibase 10.1088/1475-7516/2013/02/004}
  {\bibfield  {journal} {\bibinfo  {journal} {JCAP}\ }\textbf {\bibinfo
  {volume} {02}},\ \bibinfo {pages} {004} (\bibinfo {year} {2013})},\ \Eprint
  {http://arxiv.org/abs/1203.3542} {arXiv:1203.3542 [hep-ph]} \BibitemShut
  {NoStop}%
\bibitem [{\citenamefont {Hoferichter}\ \emph {et~al.}(2019)\citenamefont
  {Hoferichter}, \citenamefont {Klos}, \citenamefont {Men\'endez},\ and\
  \citenamefont {Schwenk}}]{Hoferichter:2018acd}%
  \BibitemOpen
  \bibfield  {author} {\bibinfo {author} {\bibfnamefont {Martin}\ \bibnamefont
  {Hoferichter}}, \bibinfo {author} {\bibfnamefont {Philipp}\ \bibnamefont
  {Klos}}, \bibinfo {author} {\bibfnamefont {Javier}\ \bibnamefont
  {Men\'endez}}, \ and\ \bibinfo {author} {\bibfnamefont {Achim}\ \bibnamefont
  {Schwenk}},\ }\bibfield  {title} {\enquote {\bibinfo {title} {{Nuclear
  structure factors for general spin-independent WIMP-nucleus scattering}},}\
  }\href {\doibase 10.1103/PhysRevD.99.055031} {\bibfield  {journal} {\bibinfo
  {journal} {Phys. Rev. D}\ }\textbf {\bibinfo {volume} {99}},\ \bibinfo
  {pages} {055031} (\bibinfo {year} {2019})},\ \Eprint
  {http://arxiv.org/abs/1812.05617} {arXiv:1812.05617 [hep-ph]} \BibitemShut
  {NoStop}%
\bibitem [{\citenamefont {Amar\'e}\ \emph {et~al.}(2019)\citenamefont {Amar\'e}
  \emph {et~al.}}]{ANAIS:2019jul}%
  \BibitemOpen
  \bibfield  {author} {\bibinfo {author} {\bibfnamefont {J.}~\bibnamefont
  {Amar\'e}} \emph {et~al.},\ }\bibfield  {title} {\enquote {\bibinfo {title}
  {{First Results on Dark Matter Annual Modulation from the ANAIS-112
  Experiment}},}\ }\href {\doibase 10.1103/PhysRevLett.123.031301} {\bibfield
  {journal} {\bibinfo  {journal} {Phys. Rev. Lett.}\ }\textbf {\bibinfo
  {volume} {123}},\ \bibinfo {pages} {031301} (\bibinfo {year} {2019})},\
  \Eprint {http://arxiv.org/abs/1903.03973} {arXiv:1903.03973 [astro-ph.IM]}
  \BibitemShut {NoStop}%
\bibitem [{\citenamefont {Agnese}\ \emph {et~al.}(2019)\citenamefont {Agnese}
  \emph {et~al.}}]{SuperCDMS:2018PL}%
  \BibitemOpen
  \bibfield  {author} {\bibinfo {author} {\bibfnamefont {R.}~\bibnamefont
  {Agnese}} \emph {et~al.} (\bibinfo {collaboration} {SuperCDMS}),\ }\bibfield
  {title} {\enquote {\bibinfo {title} {{Search for Low-Mass Dark Matter with
  CDMSlite Using a Profile Likelihood Fit}},}\ }\href {\doibase
  10.1103/PhysRevD.99.062001} {\bibfield  {journal} {\bibinfo  {journal} {Phys.
  Rev. D}\ }\textbf {\bibinfo {volume} {99}},\ \bibinfo {pages} {062001}
  (\bibinfo {year} {2019})},\ \Eprint {http://arxiv.org/abs/1808.09098}
  {arXiv:1808.09098 [astro-ph.CO]} \BibitemShut {NoStop}%
\bibitem [{\citenamefont {Aguilar-Arevalo}\ \emph {et~al.}(2024)\citenamefont
  {Aguilar-Arevalo} \emph {et~al.}}]{DAMIC:2023ela}%
  \BibitemOpen
  \bibfield  {author} {\bibinfo {author} {\bibfnamefont {A.}~\bibnamefont
  {Aguilar-Arevalo}} \emph {et~al.} (\bibinfo {collaboration} {DAMIC, DAMIC-M,
  SENSEI}),\ }\bibfield  {title} {\enquote {\bibinfo {title} {{Confirmation of
  the spectral excess in DAMIC at SNOLAB with skipper CCDs}},}\ }\href
  {\doibase 10.1103/PhysRevD.109.062007} {\bibfield  {journal} {\bibinfo
  {journal} {Phys. Rev. D}\ }\textbf {\bibinfo {volume} {109}},\ \bibinfo
  {pages} {062007} (\bibinfo {year} {2024})},\ \Eprint
  {http://arxiv.org/abs/2306.01717} {arXiv:2306.01717 [astro-ph.CO]}
  \BibitemShut {NoStop}%
\bibitem [{\citenamefont {Aguilar-Arevalo}\ \emph {et~al.}(2020)\citenamefont
  {Aguilar-Arevalo} \emph {et~al.}}]{DAMIC:2020prl}%
  \BibitemOpen
  \bibfield  {author} {\bibinfo {author} {\bibfnamefont {A.}~\bibnamefont
  {Aguilar-Arevalo}} \emph {et~al.} (\bibinfo {collaboration} {DAMIC}),\
  }\bibfield  {title} {\enquote {\bibinfo {title} {{Results on low-mass weakly
  interacting massive particles from a 11 kg-day target exposure of DAMIC at
  SNOLAB}},}\ }\href {\doibase 10.1103/PhysRevLett.125.241803} {\bibfield
  {journal} {\bibinfo  {journal} {Phys. Rev. Lett.}\ }\textbf {\bibinfo
  {volume} {125}},\ \bibinfo {pages} {241803} (\bibinfo {year} {2020})},\
  \Eprint {http://arxiv.org/abs/2007.15622} {arXiv:2007.15622 [astro-ph.CO]}
  \BibitemShut {NoStop}%
\bibitem [{\citenamefont {Aguilar-Arevalo}\ \emph {et~al.}(2022)\citenamefont
  {Aguilar-Arevalo} \emph {et~al.}}]{DAMIC:2021prd}%
  \BibitemOpen
  \bibfield  {author} {\bibinfo {author} {\bibfnamefont {A.}~\bibnamefont
  {Aguilar-Arevalo}} \emph {et~al.} (\bibinfo {collaboration} {DAMIC}),\
  }\bibfield  {title} {\enquote {\bibinfo {title} {{Characterization of the
  background spectrum in DAMIC at SNOLAB}},}\ }\href {\doibase
  10.1103/PhysRevD.105.062003} {\bibfield  {journal} {\bibinfo  {journal}
  {Phys. Rev. D}\ }\textbf {\bibinfo {volume} {105}},\ \bibinfo {pages}
  {062003} (\bibinfo {year} {2022})},\ \Eprint
  {http://arxiv.org/abs/2110.13133} {arXiv:2110.13133 [hep-ex]} \BibitemShut
  {NoStop}%
\bibitem [{sno(2006)}]{snolabuh}%
  \BibitemOpen
  \href@noop {} {\emph {\bibinfo {title} {{SNOLAB User's Handbook}}}},\
  \bibinfo {number} {Rev. 2}\ (\bibinfo {year} {2006})\ p.~\bibinfo {pages}
  {13}\BibitemShut {NoStop}%
\bibitem [{\citenamefont {Lewin}\ and\ \citenamefont
  {Smith}(1996)}]{Lewin:1995rx}%
  \BibitemOpen
  \bibfield  {author} {\bibinfo {author} {\bibfnamefont {J.~D.}\ \bibnamefont
  {Lewin}}\ and\ \bibinfo {author} {\bibfnamefont {P.~F.}\ \bibnamefont
  {Smith}},\ }\bibfield  {title} {\enquote {\bibinfo {title} {{Review of
  mathematics, numerical factors, and corrections for dark matter experiments
  based on elastic nuclear recoil}},}\ }\href {\doibase
  10.1016/S0927-6505(96)00047-3} {\bibfield  {journal} {\bibinfo  {journal}
  {Astropart. Phys.}\ }\textbf {\bibinfo {volume} {6}},\ \bibinfo {pages}
  {87--112} (\bibinfo {year} {1996})}\BibitemShut {NoStop}%
\bibitem [{\citenamefont {Baxter}\ \emph {et~al.}(2021)\citenamefont {Baxter}
  \emph {et~al.}}]{Baxter:2021pqo}%
  \BibitemOpen
  \bibfield  {author} {\bibinfo {author} {\bibfnamefont {D.}~\bibnamefont
  {Baxter}} \emph {et~al.},\ }\bibfield  {title} {\enquote {\bibinfo {title}
  {{Recommended conventions for reporting results from direct dark matter
  searches}},}\ }\href {\doibase 10.1140/epjc/s10052-021-09655-y} {\bibfield
  {journal} {\bibinfo  {journal} {Eur. Phys. J. C}\ }\textbf {\bibinfo {volume}
  {81}},\ \bibinfo {pages} {907} (\bibinfo {year} {2021})},\ \Eprint
  {http://arxiv.org/abs/2105.00599} {arXiv:2105.00599 [hep-ex]} \BibitemShut
  {NoStop}%
\bibitem [{\citenamefont {Chavarria}\ \emph {et~al.}(2016)\citenamefont
  {Chavarria} \emph {et~al.}}]{Chavarria:2016xsi}%
  \BibitemOpen
  \bibfield  {author} {\bibinfo {author} {\bibfnamefont {A.~E.}\ \bibnamefont
  {Chavarria}} \emph {et~al.},\ }\bibfield  {title} {\enquote {\bibinfo {title}
  {{Measurement of the ionization produced by sub-keV silicon nuclear recoils
  in a CCD dark matter detector}},}\ }\href {\doibase
  10.1103/PhysRevD.94.082007} {\bibfield  {journal} {\bibinfo  {journal} {Phys.
  Rev. D}\ }\textbf {\bibinfo {volume} {94}},\ \bibinfo {pages} {082007}
  (\bibinfo {year} {2016})},\ \Eprint {http://arxiv.org/abs/1608.00957}
  {arXiv:1608.00957 [astro-ph.IM]} \BibitemShut {NoStop}%
\bibitem [{\citenamefont {Agnese}\ \emph {et~al.}(2017)\citenamefont {Agnese}
  \emph {et~al.}}]{SuperCDMS:2016proj}%
  \BibitemOpen
  \bibfield  {author} {\bibinfo {author} {\bibfnamefont {R.}~\bibnamefont
  {Agnese}} \emph {et~al.} (\bibinfo {collaboration} {SuperCDMS}),\ }\bibfield
  {title} {\enquote {\bibinfo {title} {{Projected Sensitivity of the SuperCDMS
  SNOLAB experiment}},}\ }\href {\doibase 10.1103/PhysRevD.95.082002}
  {\bibfield  {journal} {\bibinfo  {journal} {Phys. Rev. D}\ }\textbf {\bibinfo
  {volume} {95}},\ \bibinfo {pages} {082002} (\bibinfo {year} {2017})},\
  \Eprint {http://arxiv.org/abs/1610.00006} {arXiv:1610.00006
  [physics.ins-det]} \BibitemShut {NoStop}%
\bibitem [{\citenamefont {Albakry}\ \emph {et~al.}(2023)\citenamefont {Albakry}
  \emph {et~al.}}]{SuperCDMS:2023geu}%
  \BibitemOpen
  \bibfield  {author} {\bibinfo {author} {\bibfnamefont {M.~F.}\ \bibnamefont
  {Albakry}} \emph {et~al.} (\bibinfo {collaboration} {SuperCDMS}),\ }\bibfield
   {title} {\enquote {\bibinfo {title} {{First Measurement of the
  Nuclear-Recoil Ionization Yield in Silicon at 100~eV}},}\ }\href {\doibase
  10.1103/PhysRevLett.131.091801} {\bibfield  {journal} {\bibinfo  {journal}
  {Phys. Rev. Lett.}\ }\textbf {\bibinfo {volume} {131}},\ \bibinfo {pages}
  {091801} (\bibinfo {year} {2023})},\ \Eprint
  {http://arxiv.org/abs/2303.02196} {arXiv:2303.02196 [physics.ins-det]}
  \BibitemShut {NoStop}%
\bibitem [{\citenamefont {Amole}\ \emph {et~al.}(2019)\citenamefont {Amole}
  \emph {et~al.}}]{PICO:2019vsc}%
  \BibitemOpen
  \bibfield  {author} {\bibinfo {author} {\bibfnamefont {C.}~\bibnamefont
  {Amole}} \emph {et~al.} (\bibinfo {collaboration} {PICO}),\ }\bibfield
  {title} {\enquote {\bibinfo {title} {{Dark Matter Search Results from the
  Complete Exposure of the PICO-60 C$_3$F$_8$ Bubble Chamber}},}\ }\href
  {\doibase 10.1103/PhysRevD.100.022001} {\bibfield  {journal} {\bibinfo
  {journal} {Phys. Rev. D}\ }\textbf {\bibinfo {volume} {100}},\ \bibinfo
  {pages} {022001} (\bibinfo {year} {2019})},\ \Eprint
  {http://arxiv.org/abs/1902.04031} {arXiv:1902.04031 [astro-ph.CO]}
  \BibitemShut {NoStop}%
\bibitem [{\citenamefont {Abdelhameed}\ \emph {et~al.}(2019)\citenamefont
  {Abdelhameed} \emph {et~al.}}]{CRESST:2019jnq}%
  \BibitemOpen
  \bibfield  {author} {\bibinfo {author} {\bibfnamefont {A.~H.}\ \bibnamefont
  {Abdelhameed}} \emph {et~al.} (\bibinfo {collaboration} {CRESST}),\
  }\bibfield  {title} {\enquote {\bibinfo {title} {{First results from the
  CRESST-III low-mass dark matter program}},}\ }\href {\doibase
  10.1103/PhysRevD.100.102002} {\bibfield  {journal} {\bibinfo  {journal}
  {Phys. Rev. D}\ }\textbf {\bibinfo {volume} {100}},\ \bibinfo {pages}
  {102002} (\bibinfo {year} {2019})},\ \Eprint
  {http://arxiv.org/abs/1904.00498} {arXiv:1904.00498 [astro-ph.CO]}
  \BibitemShut {NoStop}%
\bibitem [{\citenamefont {Kondev}\ \emph {et~al.}(2021)\citenamefont {Kondev},
  \citenamefont {Wang}, \citenamefont {Huang}, \citenamefont {Naimi},\ and\
  \citenamefont {Audi}}]{Kondev:2021lzi}%
  \BibitemOpen
  \bibfield  {author} {\bibinfo {author} {\bibfnamefont {F.~G.}\ \bibnamefont
  {Kondev}}, \bibinfo {author} {\bibfnamefont {M.}~\bibnamefont {Wang}},
  \bibinfo {author} {\bibfnamefont {W.~J.}\ \bibnamefont {Huang}}, \bibinfo
  {author} {\bibfnamefont {S.}~\bibnamefont {Naimi}}, \ and\ \bibinfo {author}
  {\bibfnamefont {G.}~\bibnamefont {Audi}},\ }\bibfield  {title} {\enquote
  {\bibinfo {title} {{The NUBASE2020 evaluation of nuclear physics
  properties}},}\ }\href {\doibase 10.1088/1674-1137/abddae} {\bibfield
  {journal} {\bibinfo  {journal} {Chin. Phys. C}\ }\textbf {\bibinfo {volume}
  {45}},\ \bibinfo {pages} {030001} (\bibinfo {year} {2021})}\BibitemShut
  {NoStop}%
\bibitem [{\citenamefont {Kurylov}\ and\ \citenamefont
  {Kamionkowski}(2004)}]{Kurylov:2003ra}%
  \BibitemOpen
  \bibfield  {author} {\bibinfo {author} {\bibfnamefont {Andriy}\ \bibnamefont
  {Kurylov}}\ and\ \bibinfo {author} {\bibfnamefont {Marc}\ \bibnamefont
  {Kamionkowski}},\ }\bibfield  {title} {\enquote {\bibinfo {title}
  {{Generalized analysis of weakly interacting massive particle searches}},}\
  }\href {\doibase 10.1103/PhysRevD.69.063503} {\bibfield  {journal} {\bibinfo
  {journal} {Phys. Rev. D}\ }\textbf {\bibinfo {volume} {69}},\ \bibinfo
  {pages} {063503} (\bibinfo {year} {2004})},\ \Eprint
  {http://arxiv.org/abs/hep-ph/0307185} {arXiv:hep-ph/0307185} \BibitemShut
  {NoStop}%
\bibitem [{\citenamefont {Feng}\ \emph {et~al.}(2013)\citenamefont {Feng},
  \citenamefont {Kumar}, \citenamefont {Marfatia},\ and\ \citenamefont
  {Sanford}}]{Feng:2013Snowmass}%
  \BibitemOpen
  \bibfield  {author} {\bibinfo {author} {\bibfnamefont {Jonathan~L.}\
  \bibnamefont {Feng}}, \bibinfo {author} {\bibfnamefont {Jason}\ \bibnamefont
  {Kumar}}, \bibinfo {author} {\bibfnamefont {Danny}\ \bibnamefont {Marfatia}},
  \ and\ \bibinfo {author} {\bibfnamefont {David}\ \bibnamefont {Sanford}},\
  }\bibfield  {title} {\enquote {\bibinfo {title} {{Isospin-Violating Dark
  Matter Benchmarks for Snowmass 2013}},}\ }in\ \href@noop {} {\emph {\bibinfo
  {booktitle} {{Snowmass 2013}: {Snowmass on the Mississippi}}}}\ (\bibinfo
  {year} {2013})\ \Eprint {http://arxiv.org/abs/1307.1758} {arXiv:1307.1758
  [hep-ph]} \BibitemShut {NoStop}%
\bibitem [{\citenamefont {Cirigliano}\ \emph {et~al.}(2012)\citenamefont
  {Cirigliano}, \citenamefont {Graesser},\ and\ \citenamefont
  {Ovanesyan}}]{Cirigliano:2012pq}%
  \BibitemOpen
  \bibfield  {author} {\bibinfo {author} {\bibfnamefont {Vincenzo}\
  \bibnamefont {Cirigliano}}, \bibinfo {author} {\bibfnamefont {Michael~L.}\
  \bibnamefont {Graesser}}, \ and\ \bibinfo {author} {\bibfnamefont {Grigory}\
  \bibnamefont {Ovanesyan}},\ }\bibfield  {title} {\enquote {\bibinfo {title}
  {{WIMP-nucleus scattering in chiral effective theory}},}\ }\href {\doibase
  10.1007/JHEP10(2012)025} {\bibfield  {journal} {\bibinfo  {journal} {JHEP}\
  }\textbf {\bibinfo {volume} {10}},\ \bibinfo {pages} {025} (\bibinfo {year}
  {2012})},\ \Eprint {http://arxiv.org/abs/1205.2695} {arXiv:1205.2695
  [hep-ph]} \BibitemShut {NoStop}%
\bibitem [{\citenamefont {Cirigliano}\ \emph {et~al.}(2014)\citenamefont
  {Cirigliano}, \citenamefont {Graesser}, \citenamefont {Ovanesyan},\ and\
  \citenamefont {Shoemaker}}]{Cirigliano:2013zta}%
  \BibitemOpen
  \bibfield  {author} {\bibinfo {author} {\bibfnamefont {Vincenzo}\
  \bibnamefont {Cirigliano}}, \bibinfo {author} {\bibfnamefont {Michael~L.}\
  \bibnamefont {Graesser}}, \bibinfo {author} {\bibfnamefont {Grigory}\
  \bibnamefont {Ovanesyan}}, \ and\ \bibinfo {author} {\bibfnamefont {Ian~M.}\
  \bibnamefont {Shoemaker}},\ }\bibfield  {title} {\enquote {\bibinfo {title}
  {{Shining LUX on Isospin-Violating Dark Matter Beyond Leading Order}},}\
  }\href {\doibase 10.1016/j.physletb.2014.10.058} {\bibfield  {journal}
  {\bibinfo  {journal} {Phys. Lett. B}\ }\textbf {\bibinfo {volume} {739}},\
  \bibinfo {pages} {293--301} (\bibinfo {year} {2014})},\ \Eprint
  {http://arxiv.org/abs/1311.5886} {arXiv:1311.5886 [hep-ph]} \BibitemShut
  {NoStop}%
\bibitem [{\citenamefont {Agnes}\ \emph {et~al.}(2023)\citenamefont {Agnes}
  \emph {et~al.}}]{DarkSide-50:2022qzh}%
  \BibitemOpen
  \bibfield  {author} {\bibinfo {author} {\bibfnamefont {P.}~\bibnamefont
  {Agnes}} \emph {et~al.} (\bibinfo {collaboration} {DarkSide-50}),\ }\bibfield
   {title} {\enquote {\bibinfo {title} {{Search for low-mass dark matter WIMPs
  with 12~ton-day exposure of DarkSide-50}},}\ }\href {\doibase
  10.1103/PhysRevD.107.063001} {\bibfield  {journal} {\bibinfo  {journal}
  {Phys. Rev. D}\ }\textbf {\bibinfo {volume} {107}},\ \bibinfo {pages}
  {063001} (\bibinfo {year} {2023})},\ \Eprint
  {http://arxiv.org/abs/2207.11966} {arXiv:2207.11966 [hep-ex]} \BibitemShut
  {NoStop}%
\bibitem [{\citenamefont {Akerib}\ \emph {et~al.}(2021)\citenamefont {Akerib}
  \emph {et~al.}}]{LUX:2020S2}%
  \BibitemOpen
  \bibfield  {author} {\bibinfo {author} {\bibfnamefont {D.~S.}\ \bibnamefont
  {Akerib}} \emph {et~al.} (\bibinfo {collaboration} {LUX}),\ }\bibfield
  {title} {\enquote {\bibinfo {title} {{Improving sensitivity to low-mass dark
  matter in LUX using a novel electrode background mitigation technique}},}\
  }\href {\doibase 10.1103/PhysRevD.104.012011} {\bibfield  {journal} {\bibinfo
   {journal} {Phys. Rev. D}\ }\textbf {\bibinfo {volume} {104}},\ \bibinfo
  {pages} {012011} (\bibinfo {year} {2021})},\ \Eprint
  {http://arxiv.org/abs/2011.09602} {arXiv:2011.09602 [hep-ex]} \BibitemShut
  {NoStop}%
\bibitem [{\citenamefont {Behnke}\ \emph {et~al.}(2017)\citenamefont {Behnke}
  \emph {et~al.}}]{PICASSO:2016lsk}%
  \BibitemOpen
  \bibfield  {author} {\bibinfo {author} {\bibfnamefont {E.}~\bibnamefont
  {Behnke}} \emph {et~al.},\ }\bibfield  {title} {\enquote {\bibinfo {title}
  {{Final Results of the PICASSO Dark Matter Search Experiment}},}\ }\href
  {\doibase 10.1016/j.astropartphys.2017.02.005} {\bibfield  {journal}
  {\bibinfo  {journal} {Astropart. Phys.}\ }\textbf {\bibinfo {volume} {90}},\
  \bibinfo {pages} {85--92} (\bibinfo {year} {2017})},\ \Eprint
  {http://arxiv.org/abs/1611.01499} {arXiv:1611.01499 [hep-ex]} \BibitemShut
  {NoStop}%
\bibitem [{\citenamefont {Aprile}\ \emph
  {et~al.}(2019{\natexlab{b}})\citenamefont {Aprile} \emph
  {et~al.}}]{XENON:2019migdal}%
  \BibitemOpen
  \bibfield  {author} {\bibinfo {author} {\bibfnamefont {E.}~\bibnamefont
  {Aprile}} \emph {et~al.} (\bibinfo {collaboration} {XENON}),\ }\bibfield
  {title} {\enquote {\bibinfo {title} {{Search for Light Dark Matter
  Interactions Enhanced by the Migdal Effect or Bremsstrahlung in XENON1T}},}\
  }\href {\doibase 10.1103/PhysRevLett.123.241803} {\bibfield  {journal}
  {\bibinfo  {journal} {Phys. Rev. Lett.}\ }\textbf {\bibinfo {volume} {123}},\
  \bibinfo {pages} {241803} (\bibinfo {year} {2019}{\natexlab{b}})},\ \Eprint
  {http://arxiv.org/abs/1907.12771} {arXiv:1907.12771 [hep-ex]} \BibitemShut
  {NoStop}%
\bibitem [{\citenamefont {Buckley}\ and\ \citenamefont
  {Lippincott}(2013)}]{Buckley:2013gjo}%
  \BibitemOpen
  \bibfield  {author} {\bibinfo {author} {\bibfnamefont {Matthew~R.}\
  \bibnamefont {Buckley}}\ and\ \bibinfo {author} {\bibfnamefont {W.~Hugh}\
  \bibnamefont {Lippincott}},\ }\bibfield  {title} {\enquote {\bibinfo {title}
  {{A Spin-Dependent Interpretation for Possible Signals of Light Dark
  Matter}},}\ }\href {\doibase 10.1103/PhysRevD.88.056003} {\bibfield
  {journal} {\bibinfo  {journal} {Phys. Rev. D}\ }\textbf {\bibinfo {volume}
  {88}},\ \bibinfo {pages} {056003} (\bibinfo {year} {2013})},\ \Eprint
  {http://arxiv.org/abs/1306.2349} {arXiv:1306.2349 [hep-ph]} \BibitemShut
  {NoStop}%
\bibitem [{\citenamefont {Crivellin}\ \emph {et~al.}(2014)\citenamefont
  {Crivellin}, \citenamefont {D'Eramo},\ and\ \citenamefont
  {Procura}}]{Crivellin:2014qxa}%
  \BibitemOpen
  \bibfield  {author} {\bibinfo {author} {\bibfnamefont {Andreas}\ \bibnamefont
  {Crivellin}}, \bibinfo {author} {\bibfnamefont {Francesco}\ \bibnamefont
  {D'Eramo}}, \ and\ \bibinfo {author} {\bibfnamefont {Massimiliano}\
  \bibnamefont {Procura}},\ }\bibfield  {title} {\enquote {\bibinfo {title}
  {{New Constraints on Dark Matter Effective Theories from Standard Model
  Loops}},}\ }\href {\doibase 10.1103/PhysRevLett.112.191304} {\bibfield
  {journal} {\bibinfo  {journal} {Phys. Rev. Lett.}\ }\textbf {\bibinfo
  {volume} {112}},\ \bibinfo {pages} {191304} (\bibinfo {year} {2014})},\
  \Eprint {http://arxiv.org/abs/1402.1173} {arXiv:1402.1173 [hep-ph]}
  \BibitemShut {NoStop}%
\bibitem [{\citenamefont {Bishara}\ \emph {et~al.}(2020)\citenamefont
  {Bishara}, \citenamefont {Brod}, \citenamefont {Grinstein},\ and\
  \citenamefont {Zupan}}]{Bishara:2018vix}%
  \BibitemOpen
  \bibfield  {author} {\bibinfo {author} {\bibfnamefont {Fady}\ \bibnamefont
  {Bishara}}, \bibinfo {author} {\bibfnamefont {Joachim}\ \bibnamefont {Brod}},
  \bibinfo {author} {\bibfnamefont {Benjamin}\ \bibnamefont {Grinstein}}, \
  and\ \bibinfo {author} {\bibfnamefont {Jure}\ \bibnamefont {Zupan}},\
  }\bibfield  {title} {\enquote {\bibinfo {title} {{Renormalization Group
  Effects in Dark Matter Interactions}},}\ }\href {\doibase
  10.1007/JHEP03(2020)089} {\bibfield  {journal} {\bibinfo  {journal} {JHEP}\
  }\textbf {\bibinfo {volume} {03}},\ \bibinfo {pages} {089} (\bibinfo {year}
  {2020})},\ \Eprint {http://arxiv.org/abs/1809.03506} {arXiv:1809.03506
  [hep-ph]} \BibitemShut {NoStop}%
\bibitem [{\citenamefont {Cheek}\ \emph {et~al.}(2023)\citenamefont {Cheek},
  \citenamefont {Price},\ and\ \citenamefont {Sandford}}]{Cheek:2023zhv}%
  \BibitemOpen
  \bibfield  {author} {\bibinfo {author} {\bibfnamefont {Andrew}\ \bibnamefont
  {Cheek}}, \bibinfo {author} {\bibfnamefont {Darren~D.}\ \bibnamefont
  {Price}}, \ and\ \bibinfo {author} {\bibfnamefont {Ellen~M.}\ \bibnamefont
  {Sandford}},\ }\bibfield  {title} {\enquote {\bibinfo {title}
  {{Isospin-violating dark matter at liquid noble detectors: new constraints,
  future projections, and an exploration of target complementarity}},}\ }\href
  {\doibase 10.1140/epjc/s10052-023-11826-y} {\bibfield  {journal} {\bibinfo
  {journal} {Eur. Phys. J. C}\ }\textbf {\bibinfo {volume} {83}},\ \bibinfo
  {pages} {914} (\bibinfo {year} {2023})},\ \Eprint
  {http://arxiv.org/abs/2302.05458} {arXiv:2302.05458 [hep-ph]} \BibitemShut
  {NoStop}%
\bibitem [{\citenamefont {Agnese}\ \emph {et~al.}(2018)\citenamefont {Agnese}
  \emph {et~al.}}]{SuperCDMS:2017SD}%
  \BibitemOpen
  \bibfield  {author} {\bibinfo {author} {\bibfnamefont {R.}~\bibnamefont
  {Agnese}} \emph {et~al.} (\bibinfo {collaboration} {SuperCDMS}),\ }\bibfield
  {title} {\enquote {\bibinfo {title} {{Low-mass dark matter search with
  CDMSlite}},}\ }\href {\doibase 10.1103/PhysRevD.97.022002} {\bibfield
  {journal} {\bibinfo  {journal} {Phys. Rev. D}\ }\textbf {\bibinfo {volume}
  {97}},\ \bibinfo {pages} {022002} (\bibinfo {year} {2018})},\ \Eprint
  {http://arxiv.org/abs/1707.01632} {arXiv:1707.01632 [astro-ph.CO]}
  \BibitemShut {NoStop}%
\bibitem [{\citenamefont {Ibe}\ \emph {et~al.}(2018)\citenamefont {Ibe},
  \citenamefont {Nakano}, \citenamefont {Shoji},\ and\ \citenamefont
  {Suzuki}}]{Ibe:2017yqa}%
  \BibitemOpen
  \bibfield  {author} {\bibinfo {author} {\bibfnamefont {Masahiro}\
  \bibnamefont {Ibe}}, \bibinfo {author} {\bibfnamefont {Wakutaka}\
  \bibnamefont {Nakano}}, \bibinfo {author} {\bibfnamefont {Yutaro}\
  \bibnamefont {Shoji}}, \ and\ \bibinfo {author} {\bibfnamefont {Kazumine}\
  \bibnamefont {Suzuki}},\ }\bibfield  {title} {\enquote {\bibinfo {title}
  {{Migdal Effect in Dark Matter Direct Detection Experiments}},}\ }\href
  {\doibase 10.1007/JHEP03(2018)194} {\bibfield  {journal} {\bibinfo  {journal}
  {JHEP}\ }\textbf {\bibinfo {volume} {03}},\ \bibinfo {pages} {194} (\bibinfo
  {year} {2018})},\ \Eprint {http://arxiv.org/abs/1707.07258} {arXiv:1707.07258
  [hep-ph]} \BibitemShut {NoStop}%
\bibitem [{\citenamefont {Xu}\ \emph {et~al.}(2024)\citenamefont {Xu} \emph
  {et~al.}}]{Xu:2023wev}%
  \BibitemOpen
  \bibfield  {author} {\bibinfo {author} {\bibfnamefont {Jingke}\ \bibnamefont
  {Xu}} \emph {et~al.},\ }\bibfield  {title} {\enquote {\bibinfo {title}
  {{Search for the Migdal effect in liquid xenon with keV-level nuclear
  recoils}},}\ }\href {\doibase 10.1103/PhysRevD.109.L051101} {\bibfield
  {journal} {\bibinfo  {journal} {Phys. Rev. D}\ }\textbf {\bibinfo {volume}
  {109}},\ \bibinfo {pages} {L051101} (\bibinfo {year} {2024})},\ \Eprint
  {http://arxiv.org/abs/2307.12952} {arXiv:2307.12952 [hep-ex]} \BibitemShut
  {NoStop}%
\bibitem [{\citenamefont {Privitera}(2024)}]{Privitera:2024tpq}%
  \BibitemOpen
  \bibfield  {author} {\bibinfo {author} {\bibfnamefont {Paolo}\ \bibnamefont
  {Privitera}} (\bibinfo {collaboration} {DAMIC-M}),\ }\bibfield  {title}
  {\enquote {\bibinfo {title} {{The DAMIC-M experiment: status and first
  results}},}\ }\href {\doibase 10.22323/1.441.0066} {\bibfield  {journal}
  {\bibinfo  {journal} {PoS}\ }\textbf {\bibinfo {volume} {TAUP2023}},\
  \bibinfo {pages} {066} (\bibinfo {year} {2024})}\BibitemShut {NoStop}%
\bibitem [{\citenamefont {Barak}\ \emph {et~al.}(2020)\citenamefont {Barak}
  \emph {et~al.}}]{SENSEI:2020dpa}%
  \BibitemOpen
  \bibfield  {author} {\bibinfo {author} {\bibfnamefont {Liron}\ \bibnamefont
  {Barak}} \emph {et~al.} (\bibinfo {collaboration} {SENSEI}),\ }\bibfield
  {title} {\enquote {\bibinfo {title} {{SENSEI: Direct-Detection Results on
  sub-GeV Dark Matter from a New Skipper-CCD}},}\ }\href {\doibase
  10.1103/PhysRevLett.125.171802} {\bibfield  {journal} {\bibinfo  {journal}
  {Phys. Rev. Lett.}\ }\textbf {\bibinfo {volume} {125}},\ \bibinfo {pages}
  {171802} (\bibinfo {year} {2020})},\ \Eprint
  {http://arxiv.org/abs/2004.11378} {arXiv:2004.11378 [astro-ph.CO]}
  \BibitemShut {NoStop}%
\bibitem [{\citenamefont {Albakry}\ \emph {et~al.}(2022)\citenamefont {Albakry}
  \emph {et~al.}}]{SuperCDMS:2022kse}%
  \BibitemOpen
  \bibfield  {author} {\bibinfo {author} {\bibfnamefont {M.~F.}\ \bibnamefont
  {Albakry}} \emph {et~al.} (\bibinfo {collaboration} {SuperCDMS}),\ }\bibfield
   {title} {\enquote {\bibinfo {title} {{A Strategy for Low-Mass Dark Matter
  Searches with Cryogenic Detectors in the SuperCDMS SNOLAB Facility}},}\ }in\
  \href@noop {} {\emph {\bibinfo {booktitle} {{Snowmass 2021}}}}\ (\bibinfo
  {year} {2022})\ \Eprint {http://arxiv.org/abs/2203.08463} {arXiv:2203.08463
  [physics.ins-det]} \BibitemShut {NoStop}%
\end{thebibliography}%

\end{document}